\documentclass{article}

\usepackage[final,nonatbib]{ML4MH_workshop_neurips2020}
\usepackage[utf8]{inputenc} 
\usepackage[T1]{fontenc}    
\usepackage{hyperref}       
\usepackage{url}            
\usepackage{amsmath} 
\usepackage{booktabs}       
\usepackage{amsfonts}       
\usepackage{nicefrac}       
\usepackage{microtype}      
\usepackage{graphicx}  
\usepackage{bm} 
\usepackage{color} 
\usepackage{subcaption}
\usepackage{wrapfig}

\title{Passive detection of behavioral shifts\\ for suicide attempt prevention}

\author{%
Pablo Moreno-Mu\~noz$^\ast$ \hspace*{1.5cm} Lorena Romero-Medrano$^{\dagger}$ \hspace*{1.5cm}\'Angela Moreno$^{\ast\dagger}$ \\
\hspace*{0.5cm}\textbf{Jes\'us Herrera-L\'opez}$^{\ast\dagger}$\hspace*{1.15cm}\textbf{Enrique Baca-Garc\'ia}$^{\ddagger}$\hspace*{0.75cm}\textbf{Antonio Art\'es-Rodr\'iguez}$^{\ast\dagger}$\\
  $^\ast$Universidad Carlos III de Madrid, $^\ddagger$F. Jim\'enez D\'iaz Hospital, $^\dagger$Evidence-Based Behavior (eB2)\\
  \texttt{\{lorena.romero,angela.moreno,jesus.herrera\}@eb2.tech}\\
    \texttt{\{pmoreno,antonio\}@tsc.uc3m.es}, \texttt{ebaca@fjd.es}
}

\begin{document}

\maketitle

\vspace*{-0.25\baselineskip}
\begin{abstract}
	More than one million people commit suicide every year worldwide. The costs of daily cares, social stigma and treatment issues are still hard barriers to overcome in mental health. Most symptoms of mental disorders are related to the behavioral state of a patient, such as the mobility or social activity. Mobile-based technologies allow the passive collection of patients data, which supplements conventional assessments that rely on biased questionnaires and occasional medical appointments. In this work, we present a non-invasive machine learning (ML) model to detect behavioral shifts in psychiatric patients from unobtrusive data collected by a smartphone app. Our clinically validated results shed light on the idea of an early detection mobile tool for the task of suicide attempt prevention.
	
	
\end{abstract}
\vspace*{-0.25\baselineskip}
\section{Introduction}
\vspace*{-0.25\baselineskip}
Severe affective disorders with high prevalence, such as depression or bipolar diseases, are mental illnesses that affect about 2\% of the world's population \cite{james2018global,who2019}. In the worst case, more than one million people worldwide commit suicide every year. The life conditions of psychiatric patients also cause perpetual disabilities across social, labour and residential domains. The problem even worsens as their mental health diseases become chronic and sufferers are unaware of their own disability or the symptoms that are harbingers of an imminent crisis. The early detection of these relapses is one of the key milestones for suicide attempt prevention.

In practice, the degree of disability has been traditionally assessed by clinicians using periodic patient reports, structured questionnaires, the assistance of caregivers or time-consuming evaluations during periodic appointments. However, several shortcomings limit the effectiveness of these procedures, including recall bias and poor reliability of such reports due to the self-representation conditioning. 

The ubiquity of smartphone devices has motivated plenty of advances in the passive assessment of mental health patients \cite{Miller2012,osmani2015smartphones,Firth2016,Barrigon2017}. However, there is still a lack of mobile-based methods able to automatically detect abrupt behavioral shifts of clinical relevance. Having the gathered digital information from smartphones, one may build accurate behavioral representations of patients in a completely passive manner \cite{Marzano2015,Madan2010} . Such observable features, modelled from multi-source observations, are powerful objective indicators for anticipating abrupt transitions in the daily behavioral state of the patients \cite{Aledavood2015,Berrouiguet2018}.



\textbf{Contribution.} The aim of this work is to present a novel machine learning methodology for the automatic assessment of daily behavioral features as well as the early detection of behavioral instabilities from them. The smartphone-based tool is both passive and personalized, consisting of three main blocks: First, the analysis and pre-processing of app usage and mobility data for building high-dimensional representations of each patient's day. Second, the Bayesian modelling of such observations within heterogeneous likelihood functions, e.g.\ an arbitrary mix of continuous and discrete variables, for obtaining latent discrete indicators of the behavioral profiles. Third, the \textit{online} detection of change-points in the low-dimensional sequence of latent behavior identifiers. Finally, we validate the obtained results within clinical data from urgencies and hospital interventions.


\begin{figure}[!ht]
	\centering
	\includegraphics[width=0.85\textwidth]{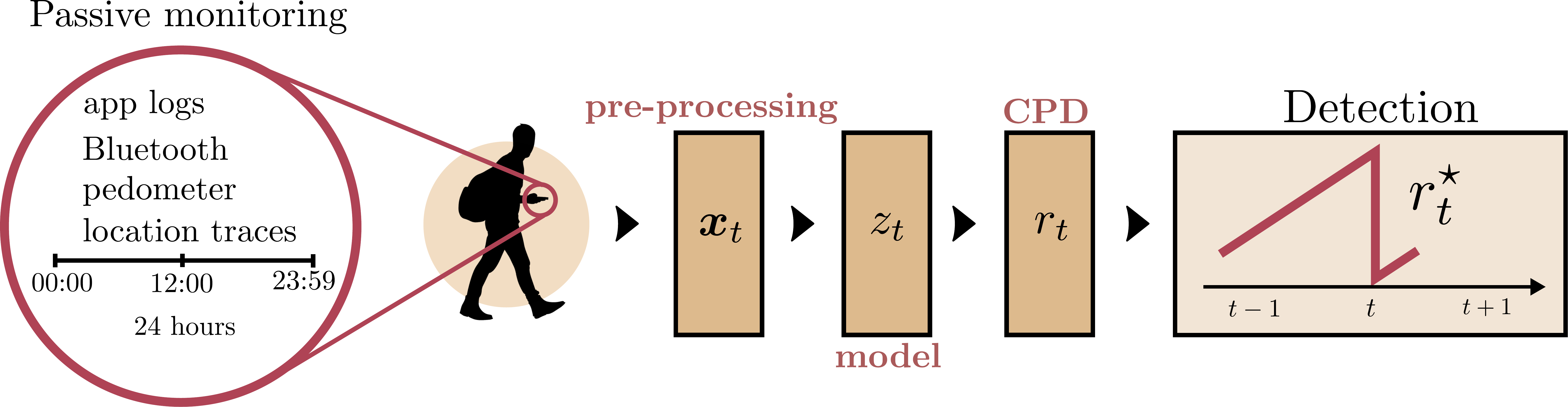}
	\caption{Illustrative diagram of the mobile-based tool for the early detection of behavioral changes. CPD is the acronym of change-point detector. The \emph{saw-tooth} red line indicates a change at time $t$.}
	\vspace*{-0.5\baselineskip}
\end{figure}
\vspace*{-0.5\baselineskip}
\section{Mobile health data}
\vspace*{-0.25\baselineskip}
Mobile electronic devices, such as personal smartphones or wearables, have a pervasive presence in our daily life. Our behavioral footprint is implicitly projected into the digital data that we generate \cite{Aledavood2015,Aledavood2015a}. We take advantage of this ubiquity, together with a monitoring \emph{app} \footnote{Evidence-Based Behavior (\url{https://eb2.tech/}).} previously used in medical studies \cite{Berrouiguet2018}, for gathering high-dimensional digital registers of patients. 


\textbf{Passive mobile sensing.} The gathering system is completely passive and unobtrusive, avoiding potential confounders due to the self-representation of the sufferer and his/her mental state. Particularly, the installed app periodically collects data (under the supervision of clinicians) from actigraphy sensors, anonymized location traces,  i.e.\ longitude-latitude, app usage logs, nearby WiFi stations and Bluetooth devices.

\textbf{Longitudinal data and daily representations.}~ 
The first stage of our work begins with the preprocessing of daily behavioral registers from the monitored patients. The sensed data is periodically stored using time slots of half-hour \cite{Eagle2009}. This precision of the data is chosen to achieve higher precision, with the cost of having high-dimensional daily vectors, e.g. $\sim$200 features. To obtain a reliable representation of one-day or time-step $t$, we choose two behavioral characteristics: mobility and social activity. For the mobility, we use the \emph{steps count} and \emph{location} traces \cite{Canzian2015}. We also compute the \emph{distance travelled} between consecutive points. Then, we generate real-valued variables, one subset per data-type, $\bm{x}^{\text{real}}\in \mathbb{R}^{D}$, that represents the total log-distance and log-steps monitored. 
$D=48$ is the number of time slots per day. For the social activity variables, such as \emph{phone usage} and \emph{home-presence indicators}, we obtain binary vectors, $\bm{x}^{\text{bin}}\in [0,1]^{D}$ where 1 indicates the mobile-phone usage and presence at home, respectively. Together, the continuous and discrete variables $\bm{x}_{t}^{\text{real}}$ and $\bm{x}_{t}^{\text{bin}}$ compose the daily and heterogeneous high-dimensional representation $\bm{x}_{t}$ of a patient. 
\vspace*{-0.25\baselineskip}
\section{Behavior modelling and change-point detection}
\vspace*{-0.25\baselineskip}
The early detection of abrupt changes in the behavioral state of a patient is an affordable challenge if we project the observations in a lower-dimensional manifold $\bm{z}$. Thus, we are interested in converting the high-dimensional sequence $\bm{x}_{1:t}$ into an interpretable discrete representation. For this task, we choose a latent class model, such that we infer the underlying sequence of discrete indicators $\bm{z}_{1:t}$ from heterogeneous observed data, e.g.\ a Gaussian-Bernoulli mixture model.

\textbf{Behavior modelling.} This stage of our work corresponds to the modelling and inference of intuitive latent categorical variables that represents the \textit{type of routine} or \textit{daily circadian profiles} of a patient. For instance, one value of the categorical latent variable may indicate a "high-activity" profile with higher level of physical activity and more time spent outdoors while another may indicate "low-activity" with less physical activity and an increase of mobile usage. To model the low-dimensional representation of the data, $z_{t}$, we use an heterogeneous circadian mixture model \cite{MorenoRamirezArtes18} for each patient. The likelihood distribution takes the form 
\begin{equation}
	p(\bm{x}_{t} \mid z_{t},\left\{\bm{\theta}_{k}^{1}, \ldots, \bm{\theta}_{k}^{M}\right\}_{k=1}^{K})=\prod_{k=1}^{K} \prod_{j=1}^{M} p(\bm{x}_{t}^{j} \mid \bm{\theta}_{k}^{j})^{\left\{1 | z_{t}=k\right\}},
	\label{eqn:likelihood}
\end{equation}
where $z_{t}\in [1,2,\cdots, K]$ indicates which component of the mixture is active in the $t$-\textit{th} day, $M$ is the number of different input data-types and $K$ is the total number of components of the mixture model. The local variables are assumed to follow
\begin{equation}
	\bm{x}_{t}^{\mathrm{real}} | z_{t} \sim \mathcal{N}\left(\mathbf{f}_{k}, \Sigma_{k}\right), \quad \bm{x}_{t}^{\text {bin }} | z_{t} \sim \operatorname{Ber}\left(\boldsymbol{\mu}_{k}\right).
\end{equation}
To infer the sequence of latent variable $\bm{z}_{1:t}$ as well as the model parameters $\{\bm{\theta}_k\}^{K}_{k=1}$and prior hyperparameters we use the \emph{expectation-maximization} (EM) algorithm. Furthermore, our presented model naturally handles partially missing observations or features \cite{Ghahramani1994,MorenoRamirezArtes18}. If one day is completely unobserved, we treat it in the following block. The concrete number of profiles $K$ for each user is selected through a method of model selection, the \emph{Bayesian information criteria} (BIC). The final goal of this stage is to obtain the posterior probabilities $p(\bm{z}_{1:t}|\bm{x}_{1:t}, \bm{\theta})$, that we will later use for the detection of behavioral change-points.

\begin{wrapfigure}[17]{r}{0.5\textwidth}
	\centering
	\vspace{-0.6cm}%
	\hspace{+0.5cm}%
	\includegraphics[width=0.48\textwidth]{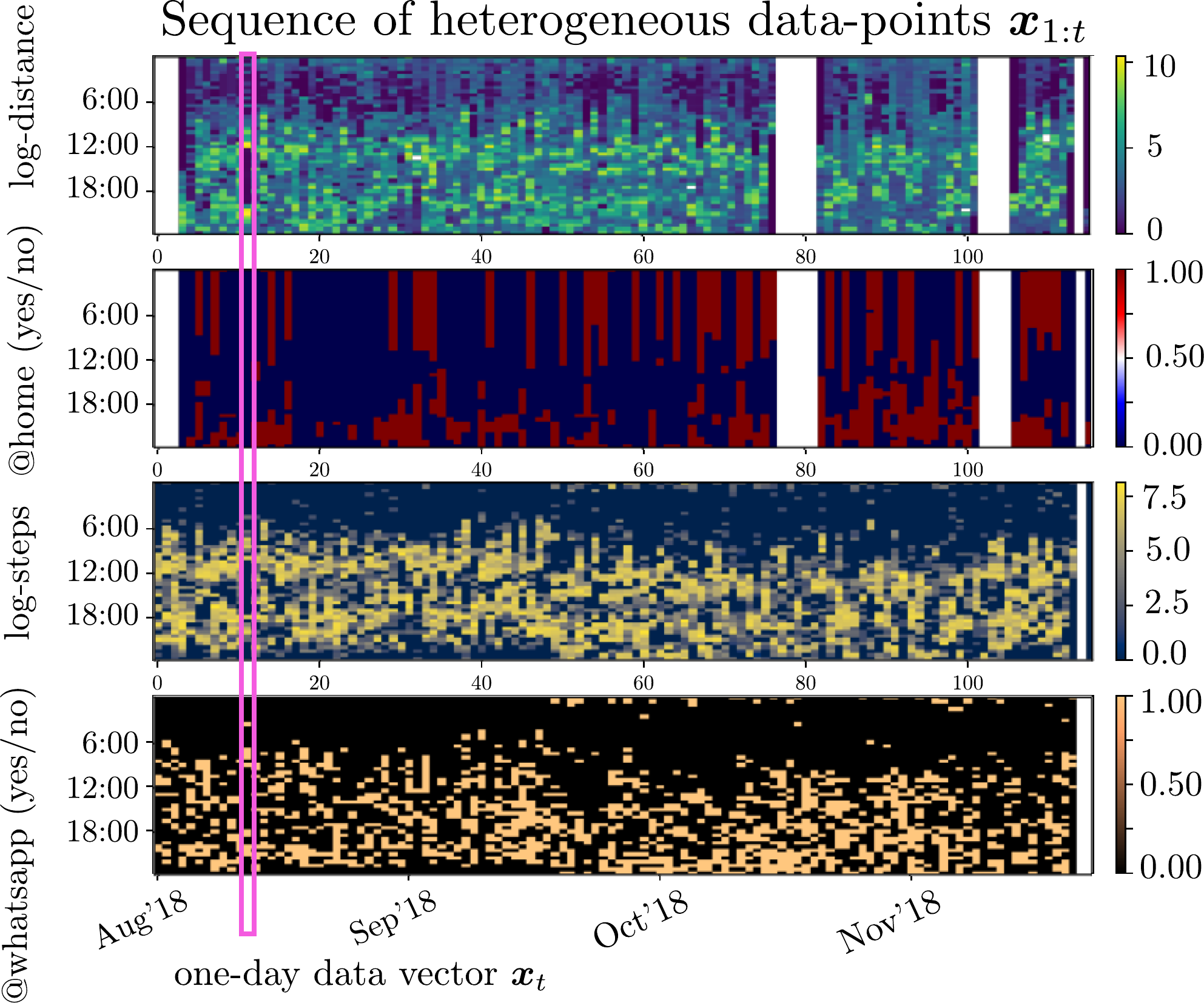} 
	\caption{Multi-sensor daily representations of the heterogeneous data gathered by smartphones.}
	\label{fig:hetero}
\end{wrapfigure}
\textbf{Change-point detection.} Given the sequence of circadian profiles $\bm{z}_{1:t}$ of a certain patient, we aim to identify abrupt transitions in their generative distribution. An example based on the previous definitions would be to think about a patient that usually has $5$ days per week of ``high-activity'' profile alternated with $2$ days of ``low-activity''. Suddenly, we observe that this proportion is inverted. In the context of chronic mental health, we must be aware of this sort of transitions as they might be warning signs of an imminent crisis. These are the change-points (CP) that we aim to detect. 


The CP detection model used was presented in \cite{romero2020multinomial}, and it is based on the hierarchical extension \cite{MorenoRamirezArtes18} of the \emph{Bayesian change-point detection} (BOCPD) algorithm \cite{Adams2007}. We assume that the main sequence of daily profiles $\bm{z}_{1:t}$ may be divided into non-overlapping partitions or subsequences, that we define as behaviors, separated by the CPs. We also assume that each behavior $\rho$ has a surrogate generative distribution $p(\bm{z}|\bm{\theta}_{\rho})$ for the profiles, observations are  i.i.d.  and the parameters $\bm{\theta}_{\rho}$ are unknown. The goal is to learn both, the unknown parameters and the CP locations.


Since we cannot observe the true sequence $\bm{z}_{1:t}$ of circadian profiles $z_t$ that represents a daily behavior, we must use the posterior distribution $p(\bm{z}_t|x_t)$ previously inferred. Based on \cite{romero2020multinomial}, we draw $S$ i.i.d. samples of the posterior distribution $z^{(1)}_t, z^{(2)}_t, \dots, z^{(S)}_t \sim p(z_{t}| x_{t}, \bm{\theta}_t)$ $\forall t$ to fully characterize the probability over the discrete latent classes. This process lead us to consider a Multinomial observation model in the CP detector, where we define the associated counting vectors $c_t \in\mathbb{Z}_+^K$. Each component $c_t^{k}:= \sum_{s=1}^S\mathbb{I}\{z_t^{(s)}=k\}$ counts the times that a  $k$-th profile has been drawn over the $S$ i.i.d.\ samples. Then, we have
\begin{equation}
	\theta_t \sim \textrm{Dirichlet}(\alpha),\qquad
	c_t \sim \textrm{Multinomial}(\theta_t, S),
\end{equation}
where $\theta_t\in\mathcal{S}^K$ and $ S\in\mathbb{N}$. The posterior of parameters has a closed-form update $\alpha' = \alpha + c_t$. The method also preserves the prior-conjugacy and is still consistent with the original BOCPD algorithm, including just one more hyperparameter in the detection model, $S$. Moreover, it has shown to increase the precision rate and to reduce the delay in the detection while keeping low computational cost.

An auxiliary discrete variable $r_t$, named the \textit{run-length} \cite{Adams2007}, is introduced into the model to carry out the inference task. It counts the number of time-steps since the last change-point occurred. The aim is to continuously infer the posterior probability $p(r_t| \bm{z}_{1:t})$ at each time step $t$, obtaining a measure of uncertainty of the last CP location given the sequence of profiles until that moment. For instance,  $p(r_{150}=5|\bm{z}_{1:150})$ would measure the probability that a change in the distribution of profiles happened $5$ days ago so, at $t = 145$. 

\textbf{Detection of behavioral shifts.} Having calculated $p(r_t| \bm{z}_{1:t})$ for every $t$,  a mechanism to define the shift between behaviors is required. We decide to use the sequence of \emph{maximum-a-posteriori} (MAP) estimates $r_t^*= \text{argmax } p(r_t| \bm{z}_{1:t})$, that are shown in red in the Fig. \ref{fig:results} and represent, for a particular $t$, the most likely day in which the current behavior started. From its definition, $r_t^*$ takes values from $0$ to $t$ and we consider that a new behavior is detected at time-step $t = t'$ if there is an abrupt decrease from $r^*_{t'-1}$ to $r^*_{t'}$. Based on experiments, we set $r^*_t \approx 0$ as the condition for the detection. 
\vspace*{-0.5\baselineskip}
\section{Experimental results}
\vspace*{-0.5\baselineskip}
In this section, we validate the accuracy of the smartphone-based ML tool for detecting behavioral changes of  mental-health patients. Our experimental results consist of three parts: i) the modelling of behavioral profiles from patients heterogeneous data with an arbitrary amount of missings, as in Fig.\ \ref{fig:hetero}, ii) the detection of change-points in the sequence of discrete behavior indicators for each patient (see Fig.\ 3) and iii) the analysis of false-alarm and sensitivity rates within the ROC curve from the timestamps provided by clinicians about true suicide attempts or inverventions in hospital urgencies. We provide the \emph{area under the ROC curve} (AUROC) metric.

\textbf{Dataset description.} A total of $301$ patients clinically diagnosed were monitored during an average period of $346$ days per patient. The data collected in the medical study \cite{berrouiguet2019combining} consist of the four registers described in Section 2 and shown in Fig.\ \ref{fig:hetero}. The total missing rate was $29\%$, with a rate of $25\%$ missings per patient.  The total number of days was $\sim104$k, and the largest register had $1492$ days.

\begin{figure}[h!]
	\centering
	\includegraphics[width=0.8\textwidth]{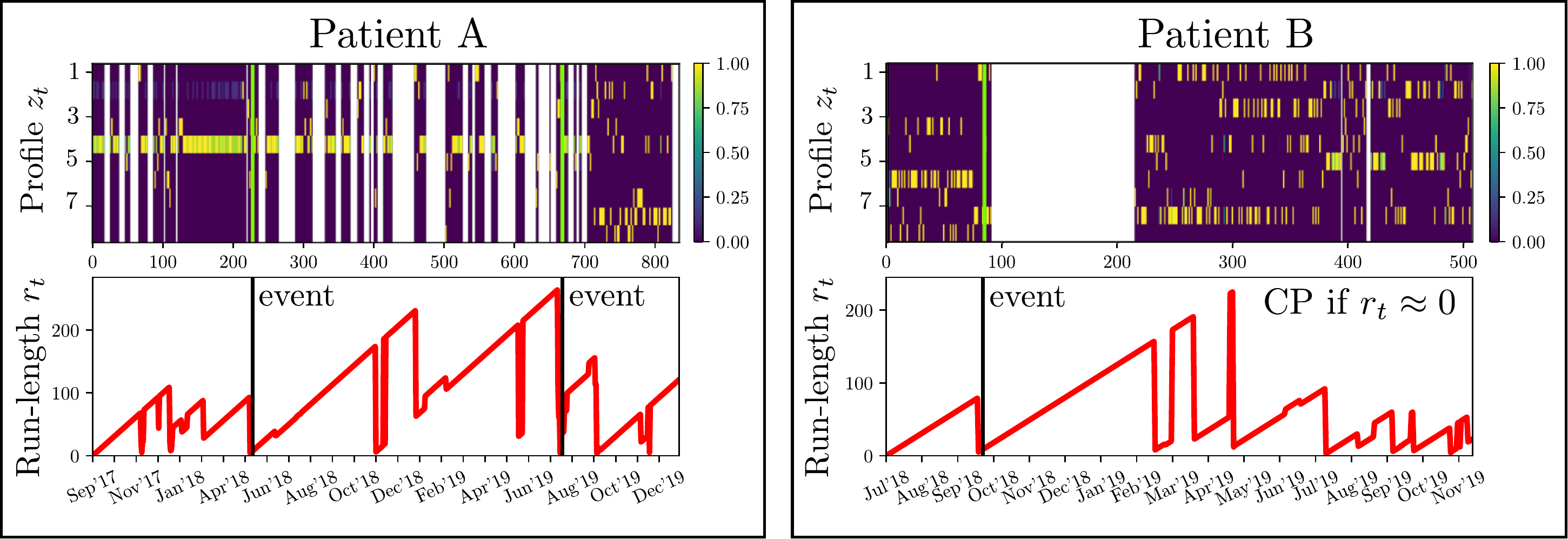} 
	\caption{Validation analysis of behavioral changes for two patients in the studied population. Vertical green and black lines indicate the \underline{\emph{true} clinical events}. Upper plots show the probabilities of each behavioral profile per day. White spaces indicate missing days. Patient A has a likely routine profile ($k=4$). Patient B turned off the smartphone after the clinical event.}
	\label{fig:results}
\end{figure}

\textbf{Clinical validation of events.} To evaluate the tool, we used the output time-steps $t$ indicated by the detector, as shown in Fig.\ \ref{fig:results}. In our case, time-steps were dates for us. We compared them with clinical time-stamps from hospital electronic health records (EHR). The dates correspond to two types of events: i) registered suicide attempts and ii) urgency interventions, i.e.\ due to crisis or self-harm. The total number of events registered by clinicians was $47$.  Only mental health patients with \emph{at least} one previous suicide attempt were considered in the medical study.


\vspace*{-0.25\baselineskip}
\vspace*{-0.25\baselineskip}

\textbf{Performance characterization metrics.} Our detector achieves an $\text{AUROC}=0.71$ in a completely passive manner. The error metric is significantly over the random guessing threshold. A key point is that the clinical scenario requires a minimum false-alarm rate, due to the psychological costs to both sufferers and caregivers. The tool can be also personalized to higher or lower probability of an event. We chose a window of one-week as the warning period, that is, a CP time-step $t$ placed six or less days before an event is considered a \emph{true positive}. This can be also tuned to have longer or shorter windows. Depending on the abruptness of the detection output (red line in Fig.\ \ref{fig:results}), clinical interventions can be delivered to patients. We remark the robustness of the early detection tool under missing observations, typically, due to smartphones are turned off or out-of-signal.

\vspace*{-0.5\baselineskip}
\section{Discussion \& future work}
\vspace*{-0.25\baselineskip}
We introduced a novel tool for the detection of behavioral shifts with application to suicide attempt prevention. Chronic mental health patients are passively monitored via their smartphones, allowing us to build high-precision representations of their mobility and social activity. We introduced a discrete latent variable model that captures the underlying behavioral profile of each day. The detection over the latent manifold is performed by an hierarchical change-point model whose robustness is guaranteed. The validated results compare the detection of events in a population of $\sim$300 mental health patients and the clinical information provided from urgencies and hospital interventions. In future work, it would be interesting to introduce other latent representations, as for instance, factorial models, highly non-linear parameterizations or disentangled variables.

\small
\bibliography{ml4mh_bibliography}

\end{document}